\documentclass[prd,showpacs,preprint,preprintnumbers]{revtex4}
\usepackage[mathscr]{eucal}
\usepackage{color}
\usepackage{amsmath}
\usepackage[dvips]{graphicx}
\usepackage{epsf}
\usepackage{bm}
\usepackage{graphicx}
\usepackage{bm}
\usepackage{color}

\newcommand{\be}{\begin{equation}}
\newcommand{\dd}{\displaystyle}
\newcommand{\ee}{\end{equation}}
\newcommand{\bea}{\begin{eqnarray}}
\newcommand{\eea}{\end{eqnarray}}

\def\dd{{\displaystyle}}

\def\slr#1{\setbox0=\hbox{$#1$}           
   \dimen0=\wd0                                 
   \setbox1=\hbox{/} \dimen1=\wd1               
   \ifdim\dimen0>\dimen1                        
      \rlap{\hbox to \dimen0{\hfil/\hfil}}      
      #1                                        
   \else                                        
      \rlap{\hbox to \dimen1{\hfil$#1$\hfil}}   
      /                                         
   \fi}

\def\be{\begin{eqnarray}}
\def\ee{\end{eqnarray}}

\begin{document}

 \preprint{BARI-TH 532/06}
\title{Chromomagnetic stability of the three flavor Larkin-Ovchinnikov-Fulde-Ferrell phase of QCD}
\date{\today}

\author{M. Ciminale}
\author{G. Nardulli}
\author{M. Ruggieri}
\affiliation{Universit\`a di Bari, I-70126 Bari, Italia}
\affiliation{I.N.F.N., Sezione di Bari, I-70126 Bari, Italia}
\author{R. Gatto}
\affiliation{D\'epart. de Physique Th\'eorique, Universit\'e de
Gen\`eve, CH-1211 Gen\`eve 4, Suisse}%
\date{\today}

\begin{abstract}
We compute in the Ginzburg-Landau approximation the gluon Meissner
masses for the  Larkin-Ovchinnikov-Fulde-Ferrell (LOFF) phase of QCD
with three flavors in the kinematical range where it is
energetically favored. We find real Meissner masses and therefore
chromomagnetic stability.

\end{abstract}


\maketitle

\section{Introduction}
The attractive interaction between quarks in the color antisymmetric
channel leads at high densities and small temperatures to their
Cooper pairing, see~\cite{barrois,Alford:1997zt} and for
reviews~\cite{Rajagopal:2000wf,Nardulli:2002ma}; in particular, for
three flavors at asymptotically high densities, the
Color-Flavor-Locking (CFL) phase is the ground state of the theory,
characterized by a spinless color- and flavor- antisymmetric diquark
condensate~\cite{Alford:1998mk}.

The physically more interesting pre-asymptotic phases, where the
mass of the strange quark  and the chemical potential differences
$\delta\mu$ due to $\beta$ equilibrium cannot be neglected, are
subject of intense study at this moment. Proposals include the 2SC
phase~\cite{Alford:1997zt}, the gapless phases
g2SC~\cite{Shovkovy:2003uu}, and gCFL
\cite{Alford:2003fq,Alford:2004hz}. Unfortunately the appearance of
imaginary gluon Meissner masses (for g2SC see~\cite{Huang:2004bg},
for gCFL see~\cite{Casalbuoni:2004tb}) makes the gapless phases
instable, and also the 2SC phase shows instability
~\cite{Huang:2004bg} (in~\cite{Huang:2005pv,Gorbar:2005rx} possible
antidotes to cure the chromo-magnetic instability are discussed).

However, for appropriate values of $\delta\mu$, quarks may form
pairs with non-vanishing total momentum: ${\bf p_1}+{\bf p_2}=2{\bf
q}\neq 0$, see~\cite{Alford:2000ze} and for a
review~\cite{Casalbuoni:2003wh}, which leads to a
Larkin-Ovchinnikov-Fulde-Ferrell (LOFF)~\cite{LOFF2} phase. For two
flavors it has been shown~\cite{Giannakis:2004pf} that the 2SC
instability implies that the LOFF phase is energetically favored;
however, it is not clear if the neutral LOFF phase can cure the
chromomagnetic instability of the two flavor superconductive quark
matter~\cite{Giannakis:2005vw,Gorbar:2005tx}.

LOFF with three flavors is however the much more difficult but
physically interesting case at intermediate densities. A first study
of the problem was carried out in a Ginzburg-Landau (GL)
approximation \cite{Casalbuoni:2005zp}. It was found that
condensation of the pairs $u-s$ and $d-u$ is possible in the form of
the inhomogeneous LOFF pairing.

The problem which comes next is then to find out whether such a phase is chromomagnetic stable. This
 is the subject of the present note. Within the GL expansion
  we find that this is indeed the case. This result is another
 indication that the LOFF phase of QCD
 plays a major role at intermediate hadronic  densities.
Since  these pre-asymptotic densities are probably relevant for the
cores of compact stars, a whole field of investigation is opened up,
with items ranging from  transport properties, cooling processes and
the glitches in the pulsar rotational frequency to the possible
implications for gravitational radiation during collisions of
compact objects and the mechanisms of gamma ray bursts. Besides,
given the universality which is peculiar to the phenomena occurring
at the Fermi surface, one expects important indications to come from
the study of the similar problems with cold fermionic
atoms~\cite{coldatomsEx}, mainly because of the greater parameter
flexibility possible in laboratory experiments.

The plan of the paper is as follows. In section~\ref{sec:review} we
discuss the model and review the results
of~\cite{Casalbuoni:2005zp}; in section~\ref{sec:GLmasses} we
present the computation of the gluon Meissner masses; in
section~\ref{s4} we discuss our results and in
section~\ref{sec:conc} we give our conclusions. The concluding
appendix contains some useful formulas.

\section{The three flavor LOFF phase of QCD}\label{sec:review}
In the CFL phase of QCD all the eight gluons acquire real Meissner
masses, while introducing the strange quark mass and the neutrality
constraints (which leads to the gCFL phase) some of these masses
become imaginary, which is a signal of instability. In this paper we
compute gluon Meissner masses in the three flavor inhomogeneous case
(LOFF phase). We will work in the GL
approximation~\cite{Casalbuoni:2005zp}. For a system of two massless
($u$ and $d$) and one massive ($s$) quarks the QCD action is:
\begin{equation}
{\cal I}=\int\!d^4x~\bar{\psi}_{i\alpha}(x)\,\left(i\,D\!\!\!\!
/^{\,\,\alpha\beta}_{\,\,ij} -M_{ij}^{\alpha\beta}+
\mu^{\alpha\beta}_{ij} \,\gamma_0\right)\,\psi_{\beta j}(x)
\label{lagr1}\ .
\end{equation}
Here $\alpha,\,\beta,\,$ are color indexes, $i,\,j\,$ flavor
indexes; $M_{ij}^{\alpha\beta} =\delta^{\alpha\beta}\, {\rm
diag}(0,0,M_s) $ is the quark mass matrix and
$D^{\alpha\beta}_{ij}=\partial_x\delta^{\alpha\beta}\delta_{ij}+
igA_aT_a^{\alpha\beta}\delta_{ij}$; $\mu_{\alpha\beta}^{ij}$ is the
matrix of the chemical potentials, see below. It   depends on $\mu$
(the average quark chemical potential), $\mu_e$ (the electron
chemical potential), and $\mu_3,\,\mu_8$, related to
color~\cite{Alford:2003fq,Buballa:2005bv}.

We treat the strange quark mass at its leading order, ie by a shift
in the strange quark chemical potential $\mu_s\to\mu_s-M_s^2/2\mu$,
and we adopt the High Density Effective Theory
(HDET)~\cite{Hong:1998tn} (see~\cite{Nardulli:2002ma} and
\cite{Schafer:2003yh} for reviews). HDET uses the fact that, for
$T\to 0$ and at weak coupling, the relevant modes of the QCD action
are those near the Fermi surface. It is useful therefore to
decompose the quark momentum as follows: $ {\bf p}=\mu{\bf n}+{\bm
\ell}$; $\mu{\bf n}$ is the large component (${\bf n}$  a unit
vector representing the quark Fermi velocity) and $\bm\ell$ is the
small residual momentum that one can take in the direction of $\bf
n$ using reparameterization invariance, so that ${\bm \ell}=\xi\bf
n$. Moreover one introduces velocity dependent fields $\psi_{\bf n}$
and $\Psi_{\bf n}$ corresponding to positive and negative energy
solutions of the Dirac equation:\begin{equation}
\psi(x)=\int\frac{d{\bf n}}{4\pi}~e^{i\,\mu{\bf n}\cdot{\bf
x}}\,\left(\psi_{\bf n}(x)+\Psi_{\bf n}(x)\right)\label{decomp} \,.
\end{equation}
 Substituting (\ref{decomp}) in
(\ref{lagr1}) and integrating out  the negative energy components,
one gets in momentum space, at the next-leading-order in $1/\mu$:
\begin{equation} {\cal L}= \,\psi_{{\bf n},i\alpha}^\dagger({\ell})\left(V \cdot
\ell_{ij}^{\alpha\beta}  + \bar\mu_{ij}^{\alpha\beta}
 + P_{\mu\nu} \left[
\frac{\ell_{\mu} \ell_{\nu}}{  \tilde V \cdot \ell + 2 \mu
}\right]_{ij}^{\alpha\beta} \right)\psi_{{\bf n},\beta j}({\ell})
\label{L11}\, ,
\end{equation}
where
$(\ell^\mu)_{ij}^{\alpha\beta}=\ell^\mu\delta_{ij}\delta^{\alpha\beta}-gA_aT_a^{\alpha\beta}
\delta_{ij}$, with $\ell^\mu=(p^0,\,\xi{\bf n})$. Moreover
$V^\mu=(1,{\bf n}),~\tilde{V}^\mu=(1,-{\bf n})$,~
$\bar\mu_{ij}^{\alpha\beta} =
\mu_{ij}^{\alpha\beta}-\mu\delta^{\alpha\beta}\delta_{ij}-
M_s^2/(2\mu)~\delta^{\alpha\beta}\delta_{ij}\delta_{i3}$ and
$P^{\mu\nu}= g^{\mu \nu} - \left(V^{\mu}\tilde V^{\nu}+ \tilde
  V^{\mu} V^{\nu} \right)/2$~.

In the color-flavor and Nambu-Gorkov (NG) basis introduced
in~\cite{Casalbuoni:2004tb} the free propagator reads
\begin{equation}
S_0~=~\left(\begin{array}{cc}
              [S_0^{11}]_{AB} & 0 \\
              0 & [S_0^{22}]_{AB}
            \end{array}
\right)~=~\delta_{AB}\left(\begin{array}{cc}
              \left(p_0-\xi+\bar\mu_A \right)^{-1} & 0 \\
              0 & \left(p_0+\xi-\bar\mu_A \right)^{-1}
            \end{array}
\right)~,\label{eq:FreeProp}
\end{equation}
where $\bar\mu_{A} \ = \ \left(\bar\mu_{ru}, \bar\mu_{gd},
\bar\mu_{bs}, \bar\mu_{rd}, \bar\mu_{gu}, \bar\mu_{rs},
\bar\mu_{bu}, \bar\mu_{gs}, \bar\mu_{bd}\right)$. Writing the
propagator as in Eq.~\eqref{eq:FreeProp} one doubles the fermion
modes, which is compensated by an extra factor $1/2$ in momentum
integration.

To keep into account quark condensation we add to the QCD action
 the bilinear quark term
\begin{equation}
{\cal I}_{\Delta}=-\frac{1}{2}\,\int\!d^4x\!\int\frac{d{\bf
n}}{4\pi}\,\Delta^{\alpha\beta}_{ij}(x)\, \,\psi^T_{\alpha i,-{\bf
n}}(x)\,C\,\gamma_5\,\psi_{j\beta,{\bf n}}(x)\,+\,h.c.
\label{gapLagrUNO}
\end{equation}
where the gap function is given by~\cite{Casalbuoni:2005zp}
\begin{equation}
\Delta^{\alpha\beta}_{ij}(x)=
\sum_{I=1}^{3}\,\Delta_I\exp\left\{2i{\bf q_I}\cdot {\bf
x}\right\}\,\epsilon^{\alpha\beta I}\,\epsilon_{ijI}~.\label{cond}
\end{equation}  Eq.
(\ref{cond}) corresponds to assume a Fulde-Ferrell ansatz for each
inhomogeneous pairing; $2{\bf q_I}$ represents the momentum of the
Cooper pair. The inverse fermion propagator in the superconductive
phase becomes
\begin{equation}
S_0^{-1}~=~\left(\begin{array}{ccc}
              [S_0^{11}]_{AB}^{-1} & & -\Delta_{AB} \\
              &&\\
              -\Delta_{AB}^* & & [S_0^{22}]_{AB}^{-1}
            \end{array}
\right)~,~~~ \Delta_{AB} =
-\sum_{I=1}^{3}\,\Delta_I\,\text{Tr}\left[F_A^T\,\epsilon_I\,F_B\,\epsilon_I\right]~.
\label{eq:InvProp}
\end{equation}

We assume in this paper that global color neutrality is reached with
almost vanishing color chemical potentials, $\mu_3 \approx \mu_8
\approx 0$. This approximation is justified because the phase
transition from the superconductive to the normal state is second
order (see below), and the color chemical potentials are suppressed
by inverse powers of $\mu$. Therefore
\begin{equation}\mu^{\alpha\beta}_{ij}=(\mu\delta_{ij}-\mu_e
Q_{ij})\delta^{\alpha\beta}=\mu_{i}\,\delta_{ij}\delta^{\alpha\beta}\
,
\end{equation} where $Q$ is the quark electric-charge matrix.

This formalism was applied in~\cite{Casalbuoni:2005zp}, where a GL
expansion of the free energy $\Omega$ was performed. We briefly
review here the results of this paper. From the ansatz~\eqref{cond}
it is clear that $\Omega$ depends on 10 parameters, ie the three
gaps, three ${\bf q}$'s and the electron chemical potential. The
energetically favored state must be a global minimum of $\Omega$ in
the space of $\Delta$'s and ${\bf q}$'s, and has to be electrically
neutral. As a consequence, one has to solve
\begin{equation}
\frac{\partial\Omega}{\partial\mu_e}=0~,~~~~~
\frac{\partial\Omega}{\partial\Delta_I}=0~,~~~~~
\frac{\partial\Omega}{\partial {\bf q}_I}=0~.
\end{equation}
The electrical neutrality condition leads to the result $\mu_e
\approx M_s^2/(4\mu)$~, which corresponds to a symmetric splitting
of the Fermi surfaces of $d$ and $s$ around the $u$ Fermi sphere. As
a consequence $\delta\mu_{du} = \delta\mu_{us} \equiv\delta\mu$ and
$\delta\mu_{ds}=2\delta\mu$. Moreover, for the computation of the
gap parameters, it is sufficient to consider the condition
$\partial\Omega/\partial {\bf q}_I=0$ at the ${\cal O}(\Delta^2)$,
which leads to the well known relation $q = 1.1997|\delta\mu|$. This
implies that $\Delta_2 = \Delta_3$ and ${\bf q}_2 = {\bf q}_3$,
which can be interpreted as the consequence of a symmetric splitting
of the Fermi surfaces, and $\Delta_1=0$, due to the larger mismatch
among the $d$ and $s$ Fermi surfaces. These results hold  in the
range $(128-150) $ MeV for the parameter $M_s^2/\mu$
\cite{Casalbuoni:2005zp}. For smaller values the LOFF free energy is
higher than the gCFL energy; for higher values the normal phase is
favored.

For this solution the GL density of free energy reads
\begin{equation}
\Omega = \Omega_n+ \frac{\alpha_2 \Delta_2^2 +
\alpha_3\,\Delta_3^2}{2} ~+~ \frac{\beta_2 \Delta_2^4 +
\beta_3\,\Delta_3^4}{4}  ~+~
\frac{\beta_{23}}{2}\Delta_2^2\Delta_3^2~,\label{eq:OmegaSimpler}
\end{equation}
where $\Omega_n$ is the usual normal contribution to the free
energy; the explicit formulas for the coefficients can be found in
the appendix. The  transition to the normal phase at
$M_s^2/\mu\approx 150$ MeV is second order, which justifies the GL
expansion.

\section{Meissner masses in Ginzburg-Landau
approximation}\label{sec:GLmasses} In the effective theory there are
two vertices describing the coupling of gluons and quarks. Either
one gluon couples to a quark and an antiquark (three-body vertex,
coupling $\sim g$) or two quarks couple to two gluons (four-body
vertex, coupling $\sim g^2$). The two vertices come from terms in
Eq.~\eqref{L11} with one or two momenta $\ell$.  At the order of
$g^2$ the four-body coupling gives rise to the contribution
$g^2\mu^2/(2\pi^2)$ identical for all the eight gluons; this result
for the LOFF phase is identical to those of the normal or the CFL
case (this term is independent of the gap parameters).The three-body
coupling gives rise to the polarization tensor: \be
i\Pi_{ab}^{\mu\nu}(x,y)=\,-\,Tr[\,i\, S(x,y)\,i\, H_{a}^{\mu}\,i\,
S(y,x)\,i\, H_{b}^{\nu}]\label{eq:Pol}\ee where the trace is over
all the internal indexes; $S(x,y)$ is the quark propagator, and
$H_{a}^{\mu}$ is the vertex matrix in the HDET formalism which can
be read from Eq.~\eqref{L11}. The quark propagator $S$ has NG
components $S^{ij}$ ($i,\,j=1,2$), see Eq.~\eqref{eq:InvProp}. At
the fourth order in $\Delta$:
\begin{gather}
S^{11}=S_{0}^{11}+S_{0}^{11}\,\Delta\,\left[S_{0}^{22}\Delta^{\star}\,
\left(S_{0}^{11}+ S_{0}^{11}\,\Delta\,S_{0}^{22}\,\Delta^{\star}\,
S_{0}^{11}\right)\right]\,,\label{s11}\\
S^{21}=S_{0}^{22}\,\Delta^{\star}\,\left(S_{0}^{11}+S_{0}^{11}\Delta
\,S_{0}^{22}\Delta^{\star}\,S_{0}^{11} \right)\,,\end{gather} where
$S_0^{ij}$ can be read in Eq.~\eqref{eq:FreeProp};
 $S^{12}$ and $S^{22}$ are obtained by the changes $11\leftrightarrow 22$ and
 $\Delta\leftrightarrow\Delta^\star$.

\begin{figure}[ht!] \centering
{\includegraphics[width=11cm]{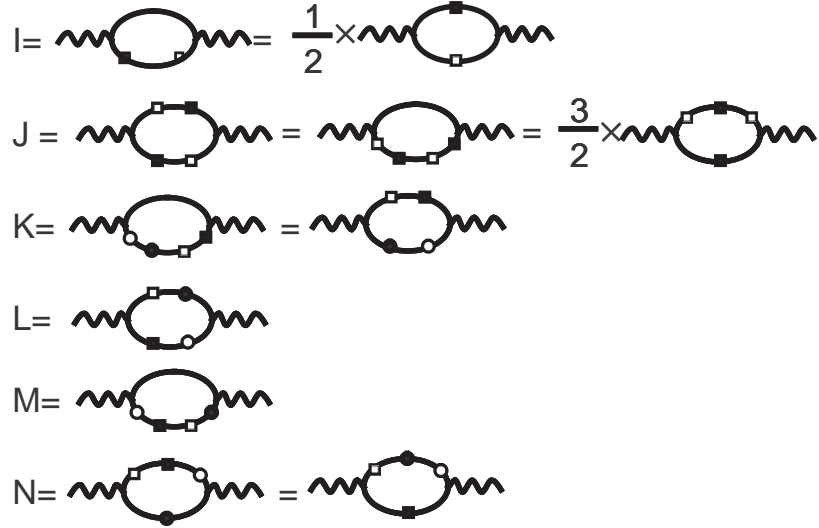}}
\caption{\label{diagrammi}{ \rm Diagrams  appearing in the GL
expansion of the gluon self-energy  in Eq.~\eqref{eq:Pol}. Wavy
lines denote gluons, solid lines are either $S_0^{11}$ and
$S_0^{22}$. Full and empty circles (squares) denote respectively the
insertion of $\Delta_2$ ($\Delta_3$) and $\Delta_2^*$
($\Delta_3^*$).  Each $\Delta^*$ is preceded by $S_0^{22}$ and
followed by $S_0^{11}$.  For a detailed description see the text.}}
\end{figure}

The different contributions  from the GL expansion are proportional
to $\Delta^0,\,\Delta^2,\,\Delta^4$. The contribution independent of
$\Delta$ is equal to $-g^2\mu^2/(2\pi^2)$ and therefore cancels out
the term  from the four-body vertex. The ${\cal O}(\Delta^2),\,{\cal
O}(\Delta^4)$ terms arise from classes of diagrams with different
topologies. Some of them are depicted in Fig.\ref{diagrammi}. The
remaining diagrams are obtained first by duplicating all the
diagrams by the exchanges
$(S_0^{11},\Delta)~\leftrightarrow~(S_0^{22},\Delta^*)$, then
duplicating the diagrams presenting both the $\Delta_2$ and the
$\Delta_3$ insertions by the exchanges
$\Delta_2\leftrightarrow\Delta_3$, and the diagrams with unequal
number of insertions on the quark lines by the exchange of the upper
and lower lines of the fermion loop.

The Meissner masses are tensors with spatial components and a
nontrivial color structure: \be\left({\cal M}^2\right)^{ij}_{ab}
\equiv - \Pi_{ab}^{ij}(p_0=0,{\bf p}=0)~,
 \ \ee
where $i,j~(a,b)$ are spatial (adjoint color) indices. We find
\begin{equation}
{\cal M}^2_{ij,11}\,=2\,I_{ij}\Delta_2^2+
\,2\,J_{ij}\Delta_{2}^4+\Delta_{3}^2\Delta_{2}^2\left(3\,K_{ij}+M_{ij}-2\,N_{ij}\right)~,\label{eq:M11}
\end{equation}
\begin{equation}
{\cal M}^2_{ij,66}\,=2\,I_{ij}(\Delta_2^2 +
\Delta_3^2)+\,2J_{ij}(\Delta_{2}^4+\Delta_{3}^4)
+\Delta_{3}^2\Delta_{2}^2\left(K_{ij}+2\,M_{ij}\right)~,\label{eq:M66}
\end{equation}
\begin{equation}
{\cal M}^2_{ij,33}\,=2\,I_{ij}\Delta_2^2+\,3\,J_{ij}\Delta_{2}^4
+\Delta_{3}^2\Delta_{2}^2\left(3\,K_{ij}+M_{ij}-2\,N_{ij}\right)~,\label{eq:M33}
\end{equation}
\begin{eqnarray}
&&{\cal M}^2_{ij,38}\,=-\frac{2}{\sqrt{3}}I_{ij}\Delta_2^2
\,-\sqrt{3}\,J_{ij}\Delta_{2}^4+\Delta_{3}^2\Delta_{2}^2\left(\sqrt{3}\,K_{ij}
-\frac{1}{\sqrt{3}}\,M_{ij}-\frac{2}{\sqrt{3}}\,N_{ij}+\frac{1}{\sqrt{3}}\,L_{ij}\right)~,
\nonumber\\
&&\label{eq:M38}
\end{eqnarray}
\begin{equation}
{\cal M}^2_{ij,88}
\,=I_{ij}\left(\frac{2}{3}\Delta_2^2+\frac{8}{3}\Delta_3^2\right)\,+
\,J_{ij}(\Delta_{2}^4+4\Delta_{3}^4)+\Delta_{3}^2\Delta_{2}^2\left(K_{ij}
+\frac{5}{3}\,M_{ij}-\frac{2}{3}\,N_{ij}-\frac{2}{3}\,L_{ij}\right)
\label{eq:M88}
\end{equation}
(the expressions of $I$, $J$, $K$, $L$, $M$ and $N$, represented in
Fig.~\ref{diagrammi}, are given in the appendix); moreover the
following identities hold: ${\cal M}^2_{ij,22}\,=\,{\cal
M}^2_{ij,11}$~, ${\cal M}^2_{ij,77}\,=\,{\cal M}^2_{ij,66}$~, ${\cal
M}^2_{ij,44}\,=\,{\cal
M}^2_{ij,11}\left(\Delta_{2}\,\leftrightarrow\,\Delta_{3}\right)$~,
${\cal M}^2_{ij,55} \,=\,{\cal M}^2_{ij,44}$~. Since the mass matrix
of the gluons $3$ and $8$ is not diagonal in the indices of the
adjoint representation, one introduces the mass eigenstates $\tilde
A_{i3} = \cos\theta_i A_{i3} + \sin\theta_i A_{i8}$~, $\tilde A_{i8}
= -\sin\theta_i A_{i3} + \cos\theta_i A_{i8}$; the physical masses
are the corresponding eigenvalues.
\section{Results\label{s4}}
\subsection{The limit cases $\Delta_2=0$ or $\Delta_3=0$}
The limit $\Delta_2=0$ is equivalent to the two flavor LOFF phase
considered in~\cite{Giannakis:2004pf,Giannakis:2005vw}, with
\begin{equation}
\langle\psi_{\alpha i}\,C\gamma_5\,\psi_{\beta j}\rangle \propto
\Delta_3\exp\left\{2i{\bf q}_3\cdot{\bf
r}\right\}\epsilon_{\alpha\beta3}\epsilon_{ij3}~.
\end{equation}
In this limit we reproduce the results of~\cite{Giannakis:2005vw},
confirming that for small gap parameters the Meissner masses of the
screened gluons are real. First of all, for $\Delta_2 = 0$ one has
${\cal M}^2_{ij,11} = {\cal M}^2_{ij,22} = {\cal M}^2_{ij,33} =0$,
which is expected because there is an unbroken $SU(2)$ color
subgroup. Moreover ${\cal M}^2_{ij,44} = {\cal M}^2_{ij,55} = {\cal
M}^2_{ij,66} = {\cal M}^2_{ij,77}$, where ($\Delta_3=\Delta$):
\begin{eqnarray}
{\cal M}^2_{xx,44} = \frac{g^2\mu^2}{96\pi^2}\frac{\Delta^4}{(q_c^2
- \delta\mu^2)^2}~, && {\cal M}^2_{zz,44} =
\frac{g^2\mu^2}{8\pi^2}\frac{\Delta^2}{(q_c^2 - \delta\mu^2)} ~,
\end{eqnarray}
and $q_c = 1.1997\,\delta\mu$ given by $\alpha^{\prime}(q_c) = 0$.
Finally, as in Eqs. (93) and (94) of~\cite{Giannakis:2005vw},
\begin{eqnarray}
{\cal M}^2_{xx,88} = 0~, &&{\cal M}^2_{zz,88} =
\frac{g^2\mu^2}{6\pi^2}\frac{\Delta^2}{(q_c^2 - \delta\mu^2)} ~.
\end{eqnarray}
In the same way one can treat the limit $\Delta_3=0$: there are 3
massless and 5 massive gluons (with the same masses as before). The
massless particles correspond to the generators $T_4$, $T_5$,
$\tilde T=T_3/2+\sqrt {3}T_8/2$  of the unbroken $SU(2)$ subgroup.
\subsection{The three flavor LOFF case}
\begin{figure}[h!] \centering
{\includegraphics[width=8.1cm]{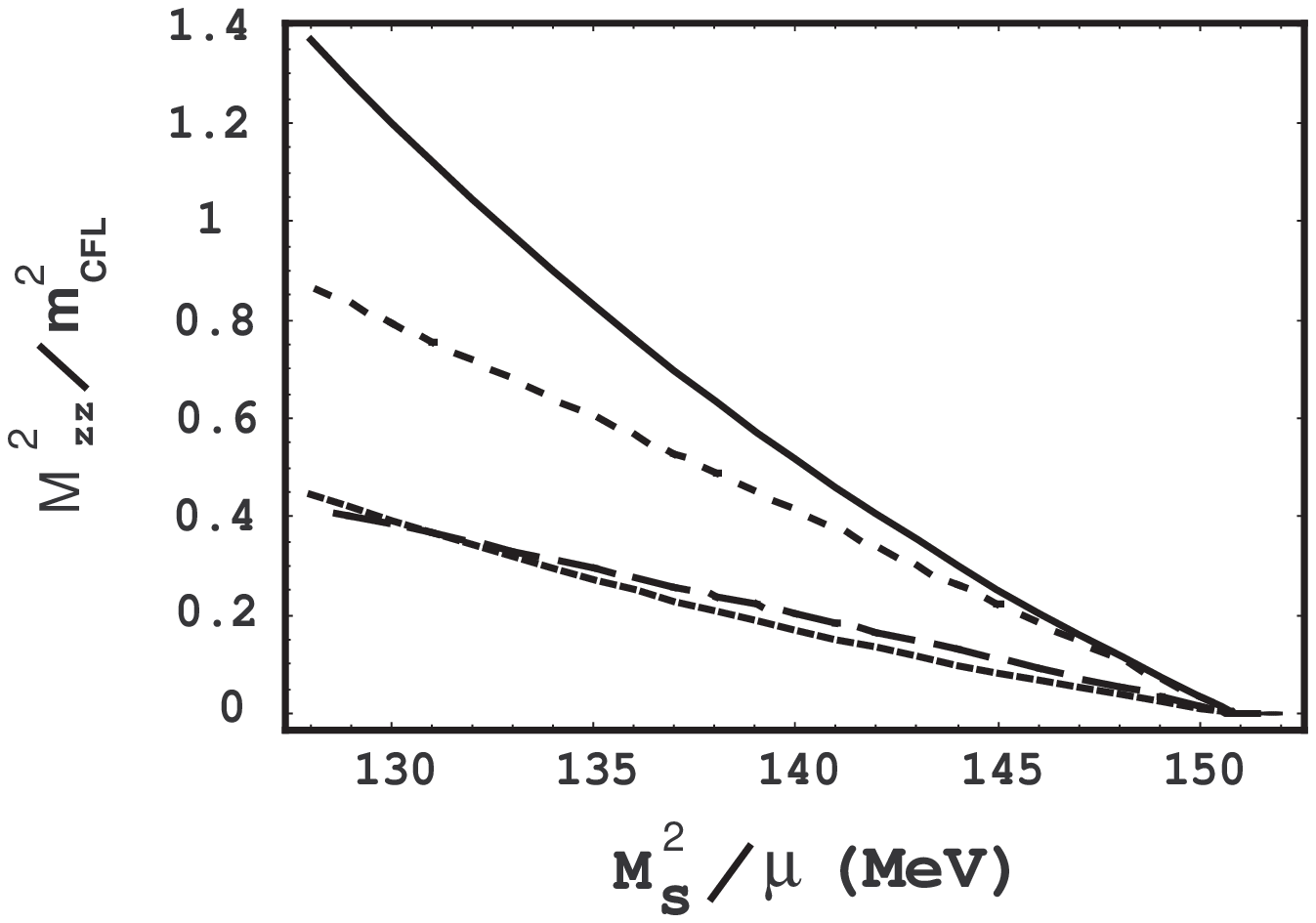}~\includegraphics[width=8.1cm]{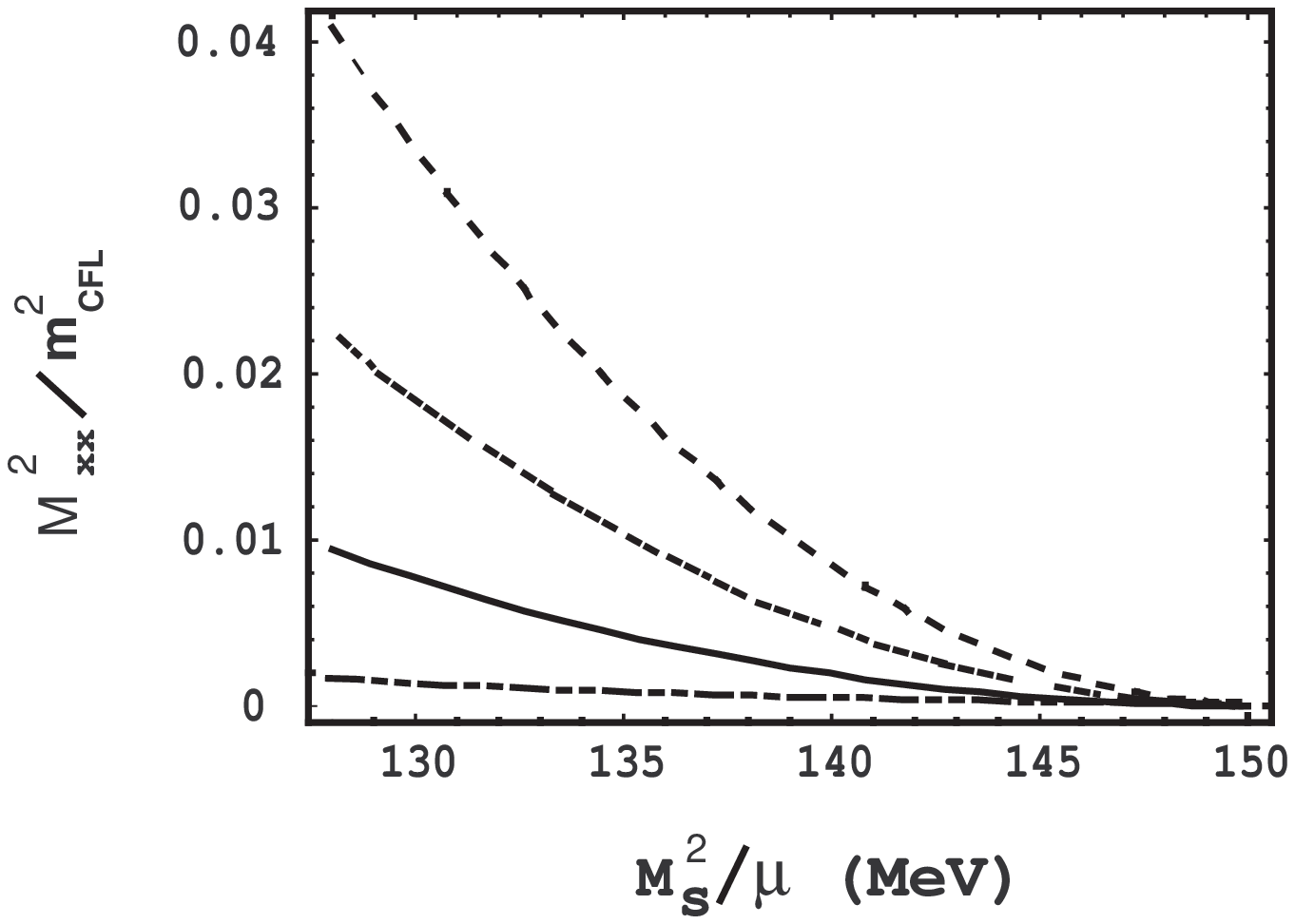}}
\caption{\label{masselong}{ \rm On the left: longitudinal squared
Meissner masses, in units of the CFL Meissner mass, vs $M_s^2 /\mu$;
from top to bottom the lines  refer to the gluons $\tilde A_3$,
$A_6$, $\tilde A_8$ (long dashed), and $A_1$ (dotted line). On the
right: transverse squared Meissner masses; from top to bottom the
lines refer to the gluons $A_6$,  $ A_1$, $\tilde A_3$, and  $\tilde
A_8$.}}
\end{figure}

In the three flavor case, for $\Delta_2=\Delta_3$, we obtain the
results shown in Fig.~\ref{masselong}. For the numerical evaluation
we have used the results of Ref.~\cite{Casalbuoni:2005zp} for the
gap $\Delta_2=\Delta_3$.

 On the left panel in
Fig.~\ref{masselong} we report the longitudinal (ie $zz$) components
of the squared Meissner masses against $M_s^2/\mu$, in units of the
CFL squared mass~\cite{Son:2000tu,Rischke:2000ra} at the ${\cal
O}(\Delta^4)$ for the representative gluons; on the right panel the
results for the transverse (ie $xx$) squared Meissner masses are
given. In both cases we obtain positive squared Meissner masses for
all the gluons. The masses not reported are obtained from those
displayed, according to the discussion at the end of section
\ref{s4}.

From the figure it is clear that for each value of the strange quark
mass, the transverse mass of a gluon is smaller than the
longitudinal one. This is so because the transverse mass is zero at
the ${\cal O}(\Delta^2)$ (see the appendix). As a consequence, it
gets contribution only from the ${\cal O}(\Delta^4)$ and is
therefore suppressed as $\Delta^2/\delta\mu^2$ in comparison with
the longitudinal one. This behavior is analogous to the two flavor
LOFF phase considered in~\cite{Giannakis:2005vw}. Moreover, we find
that the transverse mass of $\tilde A_8$, although positive, is
almost zero, being three order of magnitude smaller than the other
ones.

We conclude therefore that the LOFF phase with three flavors in the
Ginzburg-Landau limit has no chromomagnetic instability.

\section{Discussion and conclusions}\label{sec:conc}

Some final comments are in order. First of all, we have computed the
Meissner masses only for the single plane wave Fulde-Ferrell (FF)
structure, see Eq.~\eqref{cond}.
 However from the
two flavor case we know  that more complicated crystalline
structures have a lower free energy than the FF state and we expect
the same to be true in the three flavor case \cite{Rajagopal}. In
the general case one should replace~\eqref{cond} with
\begin{equation}
\Delta(x) = \sum_{i=1}^{N}\sum_{I=1}^{3}\Delta_I\exp\left\{{\bf
q}_I^i\cdot{\bf x}\right\}\epsilon_{ij I}\epsilon^{\alpha\beta I}
\end{equation}
where ${\bf q}_I^i$ $(i=1,\dots,N)$ are the momenta which define the
LOFF crystal; the geometry of the structure and the number $N$ of
plane waves should be determined by minimization of the free energy.
Once the optimal structure is found, one should compute the Meissner
masses. If this structure contains at least three linearly
independent momenta, the Meissner tensor should be positive definite
for small values of $\Delta$, since it is additive with respect to
different terms of~\eqref{cond} to order
$\Delta^2$~\cite{Giannakis:2005vw}. These considerations suggest
that a LOFF crystal can remove the chromo-magnetic instability of
the homogeneous superconductive phases of QCD, resulting as the true
vacuum of the theory. A second comment is that the results should be
extended beyond the GL expansion. Finally, in a recent
paper~\cite{Kitazawa:2006zp} it has been found that, at strong
coupling, gapless phases can be magnetic stable, and this direction
also needs to be explored.

 In conclusion  we have computed the gluon
Meissner masses in the three flavor LOFF phase of QCD using the High
Density Effective Theory and the Ginzburg-Landau approximation.
 The use of the GL expansion is
justified in the LOFF window, $128$ MeV $<M_s^2/\mu < 150$
MeV~\cite{Casalbuoni:2005zp}. In this region we find that all the
squared gluon Meissner masses are positive and therefore the LOFF
phase of three flavor QCD is free from chromomagnetic instability.

\appendix
\section{Definition of integrals}
In section \ref{sec:GLmasses} we have introduced the following
integrals:
\begin{equation}
I_{ij}=\,-\,i\frac{g^2\mu^2}{4\pi^3}\,\int\frac{
d\mathbf{n}}{4\pi}\, n_i~n_j\int\!\frac{d p_0\,
d\xi}{(p_0-\xi+\bar\mu_d)^3\cdot(p_0+\xi-\bar\mu_u+2{\bf q}\cdot{\bf
n})}~,\label{eq:Iij}
\end{equation}
\begin{equation}
J_{ij}=\,-\,i\frac{g^2\mu^2}{4\pi^3}\,\int\frac{
d\mathbf{n}}{4\pi}\, n_i~n_j\int\!\frac{d p_0\,
d\xi}{(p_0-\xi+\bar\mu_d)^4\cdot(p_0+\xi-\bar\mu_u+2{\bf q}\cdot{\bf
n})^2}~, \label{J}
\end{equation}
\begin{eqnarray}
&&K_{ij}=\,-\,i\frac{g^2\mu^2}{4\pi^3}\,\int\frac{
d\mathbf{n}}{4\pi}\, n_i~n_j\int\!\frac{d p_0\,
d\xi}{(p_0-\xi+\bar\mu_u)^4\cdot(p_0+\xi-\bar\mu_s+2{\bf q}\cdot{\bf
n})\cdot(p_0+\xi-\bar\mu_d+2{\bf q}\cdot{\bf n})}~,\nonumber\\
&&
\end{eqnarray}
\begin{equation}
L_{ij}=-\,i\frac{g^2\mu^2}{4\pi^3}\,\int\frac{ d\mathbf{n}}{4\pi}\,
 n_i~n_j\int\!\frac{d p_0\,
d\xi}{(p_0+\xi-\bar\mu_u-2{\bf q}\cdot{\bf
n})^2\cdot(p_0-\xi+\bar\mu_d)^2\cdot(p_0-\xi+\bar\mu_s)^2}~,
\end{equation}
\begin{equation}
M_{ij}=\,-\,i\frac{g^2\mu^2}{4\pi^3}\,\int\frac{
d\mathbf{n}}{4\pi}\, n_i~n_j\int\!\frac{d p_0\,
d\xi}{(p_0-\xi+\bar\mu_d)^3\cdot(p_0+\xi-\bar\mu_u+2{\bf q}\cdot{\bf
n})^2\cdot(p_0-\xi+\bar\mu_s)}~,
\end{equation}
\begin{equation}
N_{ij}=\,-\,i\frac{g^2\mu^2}{4\pi^3}\,\int\frac{
d\mathbf{n}}{4\pi}\, n_i~n_j\int\!\frac{d p_0\,
d\xi}{(p_0+\xi-\bar\mu_u+2{\bf q}\cdot{\bf
n})^3\cdot(p_0-\xi+\bar\mu_d)^2\cdot(p_0-\xi+\bar\mu_s)}~,
\label{eq:Nij}
\end{equation}
 depicted in
Fig.~\ref{diagrammi}. To evaluate these integrals first we perform a
Wick rotation $p_0 = i p_4$. The integral in $\xi$ is  computed by a
contour integration; to evaluate the integral in $p_4$ we note that
the $\xi$ integration introduces the sign factor sign$(p_4)$ and the
integrands depend only on $i p_4$, therefore one can use
\begin{equation}
\int_{-\infty}^{\infty}\!dp_4 \,\text{sign}(p_4)\,F(i p_4) = 2 i \Im
m\int_{0^+}^{\infty}\!dp_4\,F(i p_4)~;\label{eqmat}
\end{equation}
the remaining angular integrals are performed trivially. The
prescription embodied in Eq. (\ref{eqmat}) follows from the limit
$T\to 0$ once one passes from finite to zero temperature.

The integral $I_{ij} = I_{ij}(q,\delta\mu)$ has to be handled with
care. In the case of $I_{xx}$ one has
\begin{equation}
I_{xx} =
\frac{g^2}{64}\frac{1}{q}\frac{\partial\alpha(q,\delta\mu)}{\partial
q} ~,
\end{equation}
with $\alpha$ given by~\cite{Casalbuoni:2005zp}
\begin{equation}
\alpha(q,\delta\mu)=-\frac{4\mu^2}{\pi^2} \left(1 -
\frac{\delta\mu}{2q}\log\left|\frac{q+\delta\mu}{q-\delta\mu}\right|
- \frac{1}{2}\log\left|\frac{4(q^2
-\delta\mu^2)}{\Delta_0^2}\right|\right)~;
\end{equation}
the condition $\partial\Omega/\partial q = 0$ at ${\cal O}(\Delta^2)$ reads $\alpha^\prime(q)=0$: thus
$I_{xx}$ vanishes at ${\cal O}(\Delta^2)$. Defining $q_c$ the value of $q$ satisfying the equation
$\alpha^\prime(q_c)=0$ and expanding around $q_c$ one finds
\begin{equation}
I_{xx} \approx
\frac{g^2}{64}\frac{q-q_c}{q_c}\alpha^{\prime\prime}(q_c)
 ~;
\end{equation}
using $\partial \Omega/\partial q =0$ at the second order in
$\Delta^2$ one gets
\begin{equation}
\alpha^{\prime\prime}(q_c)(q-q_c) = -\frac{\Delta^2}{2}
 \left(\beta^\prime(q_c) +
 2\beta^\prime_{23}(q_c)\right)~\label{eq:AppDer}
\end{equation}
where~\cite{Casalbuoni:2005zp}
$\beta=\mu^2/[\pi^2(q^2-\delta\mu^2)]$, and \be
\beta_{23}=-\frac{2\mu^2}{\pi^2}\,\Re e\!\!\int\!\frac{d{\bf
n}}{4\pi}\,\frac{1}{(2{\bf q_3}\cdot{\bf
n}+\mu_u-\mu_d-i\epsilon)\,(2{\bf q_2}\cdot{\bf
n}+\mu_u-\mu_s-i\epsilon)}~. \ee In \eqref{eq:AppDer} the derivative
of $\beta_{23}$ is in the variable $q_3$, for fixed $q_2$; at the
end one puts $q_2 = q_3$. At the lowest non-vanishing order in
$\Delta^2$ one has
\begin{equation}
I_{xx} = -\frac{g^2}{64}\frac{\Delta^2}{2
q_c}\left(\beta^\prime(q_c) + 2\beta^\prime_{23}(q_c)\right)~.
\label{eq:IxxExp}
\end{equation}
The integral in the longitudinal case, $I_{zz}$, is treated in a
similar way,
\begin{equation}
I_{zz}(q)\approx I_{zz}(q_c) +
(q-q_c)I^\prime_{zz}(q_c)~.\label{eq:IzzExp}
\end{equation}
As a consequence, the integrals $I_{ij}$ give  contributions at the
second and  fourth order in $\Delta$.


\end{document}